\documentclass[twocolumn,preprintnumbers,floats,prd,amssymb,floatfix,nofootinbib,balancelastpage,superscriptaddress,amsmath]{revtex4-1}

\pdfoutput=1
\usepackage[utf8]{inputenc}
\usepackage{amssymb}
\usepackage{amsmath}
\usepackage{amsfonts}
\usepackage{graphicx}
\usepackage{color}
\usepackage{xspace}
\usepackage{comment}
\usepackage{hyperref}
\usepackage[normalem]{ulem}
\usepackage[section]{placeins}
\usepackage{afterpage}
\usepackage{slashed}
\usepackage{enumerate}
\usepackage{enumitem}
\usepackage{caption}
\usepackage{subfigure}
\usepackage{float}
\usepackage[T1]{fontenc}
\usepackage{mathptmx}
\usepackage{ulem}
\usepackage{enumitem}
\usepackage{verbatim}
\usepackage{multirow,rotating, booktabs}
\usepackage[dvipsnames]{xcolor}

\definecolor{dcolour}{rgb}{.5, .5, .5}
\def\gsim{\raise0.3ex\hbox{$\;>$\kern-0.75em\raise-1.1ex\hbox{$\sim\;$}}}
\def\lsim{\raise0.3ex\hbox{$\;<$\kern-0.75em\raise-1.1ex\hbox{$\sim\;$}}}
\def\gsim{\raise0.3ex\hbox{$\;>$\kern-0.75em\raise-1.1ex\hbox{$\sim\;$}}}
\def\lsim{\raise0.3ex\hbox{$\;<$\kern-0.75em\raise-1.1ex\hbox{$\sim\;$}}}

\newcommand{\ba}[1]{\begin{eqnarray} \label{(#1)}}
\newcommand{\ea}{\end{eqnarray}}

\newcommand{\mykeywords}[1]{%
  \par\vspace{0.5em}%
  \noindent\textbf{Keywords:} #1%
}

\begin{document}

\title{High-Discretization Method of Moments for Capacitance Calculation: A Cube and a Hollow Cylinder}

\author{Haiyong Gu}
\email{haiyong@xujc.com}
\affiliation{ 
School of Electronic Science and Technology, Xiamen University Tan Kah Kee College,
363105 Zhangzhou, Fujian, China}
\author{Liyuan Huang}
\email{lyhuang@xujc.com}
\affiliation{ 
School of Electronic Science and Technology, Xiamen University Tan Kah Kee College,
363105 Zhangzhou, Fujian, China}
\author{Peide Yang}
\email{ yangpd@xujc.com}
\affiliation{ 
School of Electronic Science and Technology, Xiamen University Tan Kah Kee College,
363105 Zhangzhou, Fujian, China}
\author{Tianshu Luo}
\email{ penny678@xujc.com}
\affiliation{ 
School of Electronic Science and Technology, Xiamen University Tan Kah Kee College,
363105 Zhangzhou, Fujian, China}
\author{Han Dong}
\email{ donghan@mail.nankai.edu.cn}
\affiliation{ 
School of Electronic Science and Technology, Xiamen University Tan Kah Kee College,
363105 Zhangzhou, Fujian, China}

\begin{abstract}
This paper employs the method of moments (MOM) to calculate the capacitances of a cube and a hollow cylinder. For the cube, each face was divided into a maximum of $600 \times 600$ sub-areas. By fully exploiting the geometric symmetry between sub-areas and incorporating parallel computing, computational resources were significantly conserved. Our results show that the calculated capacitance of the cube first increases and then decreases as the number of sub-areas increases. When each face was divided into $90 \times 90$ sub-areas, the capacitance of the unit cube (with an edge length of $1$ m) reached a maximum reference value of $73.519014$ pF. This indicates that higher accuracy cannot be achieved merely by indefinitely increasing the number of discretized sub-areas. Subsequently, the method was applied to compute the capacitance of a hollow cylinder. The results were compared with numerical solutions based on Lekner's theoretical formula and Cavendish’s experimental values, showing good agreement among the three.

\mykeywords{Method of Moments, Capacitance, Cube, Hollow Cylinder}
\end{abstract}
\maketitle

\section{Introduction}
\label{sec:intro}
The calculation of capacitance for a cube constitutes a classic challenge in electromagnetism, for which no analytical solution has been identified to date, and existing approaches rely exclusively on numerical methods. Polya~\cite{polya1947estimating} established the upper and lower bounds for the capacitance of a cube with edge length $a$: $0.62211a < C  < 0.71055a $ cm.  Reitan and Higgins~\cite{reitan1951calculation}  calculated its capacitance as  $ 0.6555a$ cm by dividing each face of the cube into $6 \times 6$  sub-areas. Bai and Lonngren~\cite{bai2002capacitance} employed a similar approach, dividing each face of the cube into a maximum of $20\times20$  sub-areas, and obtained a capacitance value of $0.6601a$ cm. This method was later developed into the method of moments. Sarkarati~\cite{sarkarati2022precisecalculationelectricalcapacitance} applied the Method of Moments with analytically evaluated quadruple integrals, discretizing each face into $48 \times 48$ sub-areas, and obtained a capacitance value of $0.66047a$ cm. Zhou and Szabo~\cite{zhou1994brownian} obtained a capacitance value of $0.6632a$ cm using the Brownian dynamics algorithm. Hwang~\cite{hwang2010monte} obtained a capacitance value of $0.66067813a$ cm using the refined Brownian dynamics algorithm. Read~\cite{read1997improved} calculated a capacitance value of $0.6606785a$ cm using the modified Boundary Element Method. Brown~\cite{brown1990capacity} obtained a capacitance value of $0.661a$ cm using the finite difference method. Helsing and Perfekt~\cite{helsing2012polarizabilitycapacitancecube} employed an efficient boundary integral equation solver based on kernel splitting and recursive compression, and obtained a capacitance value of $0.66067815a$ cm for the unit cube.

In Sec.~\ref{sec:cube}, we employ the method of moments to discretize each face of a cube. Our approach utilizes both even and odd partitions, and by exploiting the symmetry among the resulting sub-areas, we achieve a maximum discretization of $600\times600$  sub-areas per face. The computational efficiency is significantly enhanced through parallel computing implemented in Python code, leading to a substantial saving of computational resources. In Sec.~\ref{sec:cylinder}, we apply this method of moments to calculate the capacitance of hollow cylinders with arbitrary dimensions. Readers interested solely in the capacitance analysis of hollow cylinders may proceed directly to Sec.~\ref{sec:cylinder} without loss of continuity. The discussion and conclusions are presented in Sec.~\ref{sec:dis}. Details of this study are
listed in Apps.~\ref{appendix:ElectricPotential}-~\ref{appendix:Error}.

All models and calculations in this study use SI units. The capacitance values cited from historical literature in Sec.~\ref{sec:intro} are originally in the CGS unit of centimeters. To ensure consistency, these values were converted to farads using the relation \(1\ \text{cm (CGS)} \approx 1.1126 \times 10^{-12}\ \text{F}\). For clarity, results are primarily presented in picofarads ($1\,\mathrm{pF} = 10^{-12}\,\mathrm{F}$), with the original centimeter values provided in parentheses or captions where necessary.
\section{Capacitance of a Cube}
\label{sec:cube}
\begin{figure}[h]
    \centering
    \includegraphics[width=7cm,height=7cm]{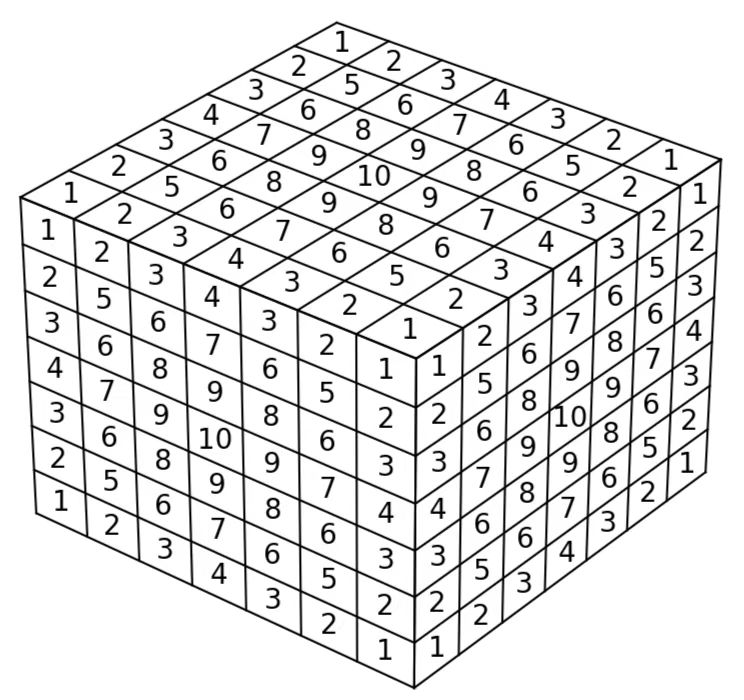}  
    \caption{Cube with Each Face Divided into $7\times7$ Sub-areas ($1$ m Side Length).
    \newline\footnotesize 
    Reproduced from [Daniel Kinseth Reitan and Thomas James Higgins. Calculation of the electrical capacitance of a cube. Journal of Applied Physics, 22(2):223–226, 1951.], with the permission of AIP Publishing.}
    \label{fig:Cube}
\end{figure}

In the method of moments, as illustrated in Fig.~\ref{fig:Cube}, each face of the cube is divided into $7\times7$  sub-areas ($1$ m side length). Due to symmetry considerations, these $49$ sub-areas are classified into ten distinct categories labeled by numbers $ 1, 2, 3 \cdots10$. Assuming that the charge carried by each sub-area is concentrated at its geometric center, the electric potential generated by a sub-area's own charge at its center can be determined through integration \eqref{eq:inter1}, yielding a value of $3.52549\sigma/(4\pi\varepsilon_0)$. The potential at this center arising from other sub-areas is approximated by $\sigma/(4\pi\varepsilon_0d)$ \eqref{eq:inter2}, where $d$ is the distance between sub-area centers and $\sigma$ is the charge density.  By computing the cumulative electric potential at the center arising from all sub-areas, we derive the following equation:
\begin{equation}
\label{eq:matrix_equation_main}
\boldsymbol{\phi} = \mathbf{P} \boldsymbol{\sigma}
\end{equation}
where
\begin{itemize}[leftmargin=0pt]
    \item $\boldsymbol{\phi} = (\phi_1, \phi_2, \dots, \phi_{10})^{\mathsf{T}}$ is the column vector of surface electric potentials,
    \item $\boldsymbol{\sigma} = (\sigma_1, \sigma_2, \dots, \sigma_{10})^{\mathsf{T}}$ is the column vector of surface charge densities, and
    \item $\mathbf{P}$ is a $10 \times 10$ coefficient matrix. The full set of matrix elements $P_{ij}$ is provided in App.~\ref{app:coefficient_matrix}.
\end{itemize}

Based on the condition of electrostatic equilibrium, which requires the electric potentials to be equal, i.e., 
$\phi_1 = \phi_2 = \phi_3 = \phi_4 = \phi_5 = \phi_6 = \phi_7 =\phi_8 = \phi_9 = \phi_{10}= \phi= 1$ V,
we solve to obtain:  

\begin{equation}
\begin{aligned}
\sigma_1 &= 3.117226765\times 10^{-12}, \quad \sigma_6 = 1.258597167\times 10^{-12}, \\
\sigma_2 &= 2.200973724\times 10^{-12}, \quad \sigma_7 = 1.228313337\times 10^{-12}, \\
\sigma_3 &= 2.059951977\times 10^{-12}, \quad \sigma_8 = 1.155954786\times 10^{-12}, \\
\sigma_4 &= 2.016716775\times 10^{-12}, \quad \sigma_9 = 1.125280939\times 10^{-12}, \\
\sigma_5 &= 1.357289047\times 10^{-12}, \quad \sigma_{10} = 1.094506278\times 10^{-12}. \\
\end{aligned}
\end{equation}

Therefore, the total charge $ Q $ can be derived:   

\begin{equation}
\begin{aligned}
Q &= 24 \sigma_1 + 48 \sigma_2 + 48 \sigma_3 + 24 \sigma_4 + 24 \sigma_5\\ 
&+ 48 \sigma_6+ 24 \sigma_7 + 24 \sigma_8 + 24 \sigma_9+ 6 \sigma_{10}\\
&= 5.11522896\times 10^{-10} \,\text{C}
\end{aligned}
\end{equation}

According to the electrostatic theorem that the capacitances of geometrically similar solids are proportional to their corresponding linear dimensions, the capacitance $ C$ of a cube with edge length $a $ m is given by:

\begin{equation}
C=\frac{Q}{\phi}\frac{a}{7} = 73.074699a \,\, \text{pF}
\end{equation}

Unlike the method of D. K. Reitan and T. J. Higgins~\cite{reitan1951calculation} , which employed only even-numbered divisions of the cube's surface (up to a maximum of $6\times6$ square sub-areas per face), our approach incorporates both even and odd divisions. Although the subsequent work by Er-Wei Bai and Karl E. Lonngren~\cite{bai2002capacitance} extended the partitioning scheme to include odd divisions (up to $20\times20$ sub-areas per face), it did not account for the geometric symmetry among the resulting sub-areas. As illustrated in Fig.~\ref{fig:Cube}, even under odd-numbered divisions, the sub-areas on the cube's surface retain inherent symmetries, allowing them to be systematically classified. In the subsequent calculations, we explicitly utilize this symmetry, which substantially reduces computational resource requirements.

\begin{figure}[h]
	\centering
	\includegraphics[width=8cm,height=8cm]{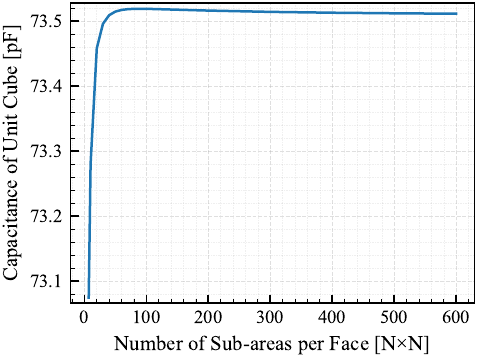}  
	\caption{Capacitance of a Unit Cube (edge length: 1
m) with Each Face Divided into N$\times$ N Sub-areas (N: X-Axis).}
	\label{fig:Cubecapacitance}
\end{figure}

We divided each face of the cube into $ 7\times7, 10\times10, 20\times20$, up to a maximum of $600\times600$ sub-areas. The calculations showed that the capacitance reached a maximum value of $73.519014a \,\, \text{pF} $ when each face was divided into $90\times90 $ sub-areas. As the number of divisions increased beyond this point, the calculated capacitance gradually decreased, as shown in Fig.~\ref{fig:Cubecapacitance} . Detailed capacitance values are provided in Table~\ref{tab:resultsB} of App.~\ref{appendix:Capacitancecube}.

The initial increase in capacitance can be explained using Thomson's theorem, which Maxwell first employed to estimate the capacitance of finite-length cylinder~\cite{cavendish1879electrical}. Thomson's theorem tells us that when a conductor reaches electrostatic equilibrium, its electrostatic field energy is minimized. The expressions for the electrostatic field energy $W$ and the total charge $Q$ can be written as follows
\begin{equation}
\begin{aligned}
W &= \frac{1}{2}\int{\phi(\xi )\sigma (\xi )dS }\\  Q &=\int{\sigma (\xi )dS } 
\end{aligned}
\end{equation}
Specifically, when the conductor reaches electrostatic equilibrium, its electrostatic field energy $W_0$ can be expressed as
\begin{equation}
{W_0} = \frac{1}{2}{\phi_0}Q = \frac{1}{2}\frac{Q}{C}Q = \frac{1}{2}\frac{{{Q^2}}}{C}
\end{equation}
Hence,
\begin{equation}
 C = \frac{{{Q^2}}}{{2{W_0}}}   
\end{equation}
According to Thomson's theorem, the electrostatic field
energy $W$ for any charge distribution satisfies $W \geq W_0$. Consequently, 
\begin{equation}
C = \frac{{{Q^2}}}{{2{W_0}}} \ge \frac{{Q^2}}{{2W}}\label{eq:c8}
\end{equation}

As the number of sub-areas increases, the charge distribution on the cube surface progressively approaches electrostatic equilibrium, i.e., $W$ approaches $W_0$ more closely, resulting in a gradual increase in capacitance.

The observed reduction in capacitance in the unit cube arises from errors due to discretization. As the number of subdivisions on each face of the cube increases, this error becomes the dominant effect. The specific and detailed reason likely stems from the potential approximation in formula \eqref{eq:inter2}. It generally overestimates the electric potential, as the potential in each sub-area is fixed at 1 V, which results in an underestimated charge density and thereby leads to a reduced capacitance. A more detailed analysis can be found in App.~\ref{appendix:Error}. This poses a challenge to the Method of Moments, as it dictates that the error introduced by discretization prevents us from obtaining the exact value of a cube's capacitance, even with an infinite increase in the number of sub-areas.

The comparison between our calculated results and those obtained by other methods is presented in Table~\ref{tab:results}. As shown in the table, when discretizing the cube surface using the method of moments, the capacitance obtained with $90\times90$ discretization is higher than that calculated with $6\times6$, $20\times20$, and $48\times48$ discretization.  In addition, the cube capacitance calculated in this paper with $90\times90$ discretization is lower than the results obtained using the Brownian dynamics algorithm~\cite{zhou1994brownian} and the finite difference method~\cite{brown1990capacity}. Since discretization errors usually lead to underestimation of capacitance, combined with the trend shown in Fig.~\ref{fig:Cubecapacitance}, the peak value obtained in this analysis—the result of $90\times90$ discretization—is selected as the reference. Therefore, it can be inferred that the actual capacitance of the cube with an edge length of $1$ m should be higher than the $90\times90$ discretization result of 73.519014 pF.

\begin{table}[h]
	\centering
    \caption{ Capacitance Values of a Unit Cube (edge length: $1$ m) Computed with Different Numerical Methods (Values in parentheses are the original data from the literature). }
\begin{tabular}{cc}
\toprule
Theoretical Method & Results $[\mathrm{pF}]$ (Original $[\mathrm{cm}]$)\\
\midrule
Method of Moments ( $6\times6$)~\cite{reitan1951calculation} & $72.930930$  ($0.6555$) \\
Method of Moments ( $20\times20$)~\cite{bai2002capacitance} & $73.442726$  ($0.6601$) \\
Method of Moments ( $48\times48$)~\cite{sarkarati2022precisecalculationelectricalcapacitance} & $73.483892$  ($0.66047$) \\
Brownian dynamics algorithm~\cite{zhou1994brownian} & $73.787632$  ($0.6632$) \\
Refined Brownian dynamics algorithm~\cite{hwang2010monte} & $73.507049$  ($0.66067813$) \\
Boundary element method~\cite{read1997improved} & $73.507980$  ($0.6606785$) \\
Finite difference method~\cite{brown1990capacity} & $73.542860$  ($0.661$) \\
Boundary integral equation solver~\cite{helsing2012polarizabilitycapacitancecube} & $73.507051$  ($0.66067815$) \\
our result ($90\times90$) & $73.519014$  \\
\bottomrule 
\end{tabular}

\label{tab:results}
\end{table}

\begin{figure}[h]
	\centering
	\includegraphics[width=9.5cm,height=7.5cm]{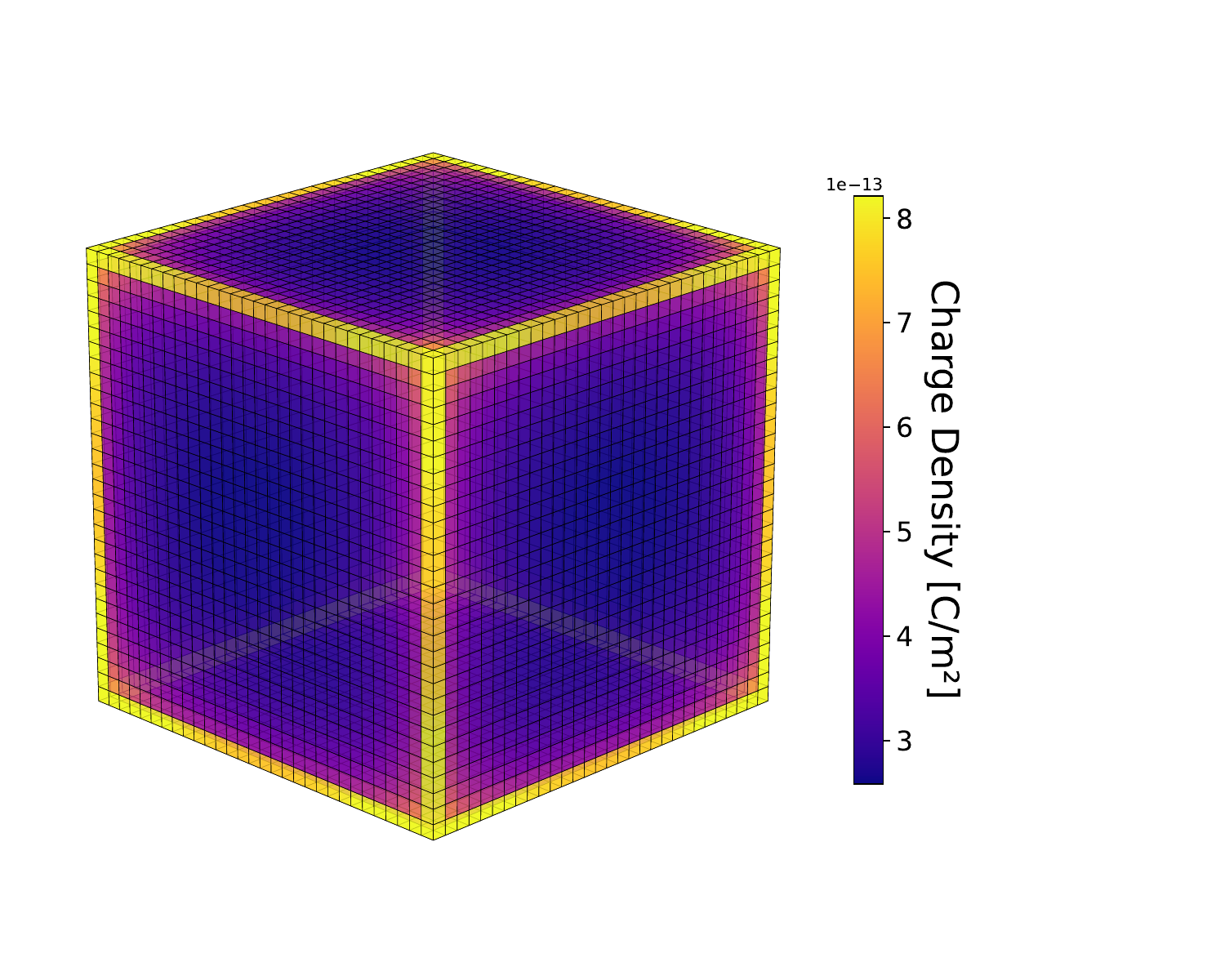}  
	\caption{Surface Charge Density Distribution on a Cube  (edge length: $1$ m)  with Each Face Divided into $30\times30$ Sub-areas.}
	\label{fig:Charge}
\end{figure}

Fig.~\ref{fig:Charge} shows the surface charge density distribution for a cube with each face divided into $30 \times 30$ sub-areas. The charge density is highest at the cube's vertices. It gradually decreases along the edges from the vertices toward the midpoints of the edges and attains its minimum at the center of each face. This
behavior is consistent with the conclusions derived from electrostatic principles.

\section{Capacitance of a Cylinder}
\label{sec:cylinder}
The capacitance of finite-length cylinders, as another classic electrostatic problem lacking an analytical solution, has been extensively investigated through numerical methods in numerous studies. Cavendish first experimentally measured the capacitance of a cylinder, and later Maxwell provided the theoretical formulation for this configuration~\cite{cavendish1879electrical}.

For a hollow cylinder of length \(L\) and radius \(R\): When \(L \gg R\), Maxwell derived the lower bound for the capacitance as \(\frac{L/2}{\ln \frac{2L}{R} - 1}\). When \(L \ll R\), the hollow cylinder reduces to a circular ring. Refs.~\cite{scharstein2007capacitance,butler1980capacitance,verolino1995capacitance,lebedev1973application} indicate that the capacitance \(C\) approaches \(\frac{\pi R}{\ln(32R/L)}\). Meanwhile, Landau~\cite{landau2013electrodynamics} calculated the capacitance of a thin ring through an elegant integral method, yielding \(\frac{\pi R }{\ln \frac{16R}{L}}\).  

For a closed cylinder, when the length is much smaller than the radius ($L \ll R$), the cylinder reduces to a disk. Maxwell's theoretical approximation $\frac{2}{\pi}\left(R + \frac{L}{4\pi}\ln\frac{2R}{L}\right) $ coincides with the capacitance of a disk derived by Landau from an ellipsoidal conductor~\cite{landau2013electrodynamics}.

Refs.~\cite{verolino1995capacitance,lebedev1973application,smythe1956charged,jackson2000charge,vainshtein1963static}  present theoretical derivations for the capacitance of the hollow cylinder with other $L/R$ ratios, while Ref.~\cite{jackson2000charge,vainshtein1963static} additionally provide the charge density distribution on the  surface of  the cylinder. For the capacitance of a hollow cylinder with arbitrary ratios of $L/R$, Paffuti~\cite{paffuti2018resultscapacitancesforcescylindrical} computed capacitances for cylinders over an extensive range of aspect ratios using a Galerkin method. Recently, de Sousa~\cite{de2026capacitance} reduced the electrostatic problem to a one-dimensional integral equation whose kernel is expressed in terms of complete elliptic integrals. A Chebyshev-weighted collocation scheme is then used, with the square-root singularity at the rims factored out analytically through the ansatz $\sigma(z) = p(x)/\sqrt{(1-x^2)}$. This produces highly accurate, extrapolated continuum-limit values for both the capacitance and the axial surface charge density over the full range of aspect ratios.  Lekner~\cite{lekner2021electrostatics} proposed the following theoretical formula:
\begin{equation}
\left| {\begin{array}{*{20}{c}}
{{K_{00}}}&{{K_{01}}}& \cdots &{{K_{0\infty }}}\\
{{K_{10}}}&{{K_{11}}}& \cdots &{{K_{1\infty }}}\\
 \vdots & \vdots & \ddots & \vdots \\
{{K_{m0}}}& \cdots & \cdots &{{K_{m\infty }}}
\end{array}} \right|\left| {\begin{array}{*{20}{c}}
{{C_0}}\\
{{C_1}}\\
 \vdots \\
{{C_\infty }}
\end{array}} \right| = \left| {\begin{array}{*{20}{c}}
1\\
0\\
 \vdots \\
0
\end{array}} \right| \label{eq:m1}
\end{equation}
Here, \( C_0 \) denotes the capacitance of the cylinder. The matrix elements \( K_{mn} \) can be expressed via the Meijer G-function~\cite{beals2013meijer}. In Mathematica notation, \( K_{mn} \) is written as:  
\begin{equation}
\begin{aligned}
K_{mn} = \frac{{(-1)^{m + n}}}{{4\pi^3\varepsilon_0 L}}   G  &\left( 
    \left[  \left[ \frac{1}{2} - m - n, \frac{1}{2} - |m - n| \right], \right. \right.\\
    & \left. \left. \left[ \frac{1}{2} + |m - n|, \frac{1}{2} + m + n \right], \right. \right. \\
    & \left. \left. \left[ \left[ 0, 0, 0 \right], \left[ 0 \right] \right], \frac{4R^2}{L^2} \right] 
    \right)
\end{aligned}
\end{equation}
Premultiplying both sides of Eq.\eqref{eq:m1} by the inverse matrix of ${K_{mn}}$ yields:  
\begin{equation}
\left| {\begin{array}{*{20}{c}}
{{C_0}}\\
{{C_1}}\\
 \vdots \\
{{C_\infty }}
\end{array}} \right| = {\left| {\begin{array}{*{20}{c}}
{{K_{00}}}&{{K_{01}}}& \cdots &{{K_{0\infty }}}\\
{{K_{10}}}&{{K_{11}}}& \cdots &{{K_{1\infty }}}\\
 \vdots & \vdots & \ddots & \vdots \\
{{K_{m0}}}& \cdots & \cdots &{{K_{m\infty }}}
\end{array}} \right|^{ - 1}}\left| {\begin{array}{*{20}{c}}
1\\
0\\
 \vdots \\
0
\end{array}} \right| 
\end{equation}
Therefore, the capacitance ${C_0}$ of the cylinder can be derived
\begin{equation}
{C_0} = {L_{00}}\label{eq:r1}
\end{equation}
${L_{00}}$ is the first matrix element of the inverse matrix of ${K_{mn}}$. The capacitance for a hollow cylinder with any aspect ratio \( L/R \)  can then be computed by truncating this infinite matrix. If the reader is interested in a more detailed derivation, they are referred to Ref.~\cite{lekner2021electrostatics}. 

\begin{figure}[h]
	\centering
	\includegraphics[width=8cm,height=4cm]{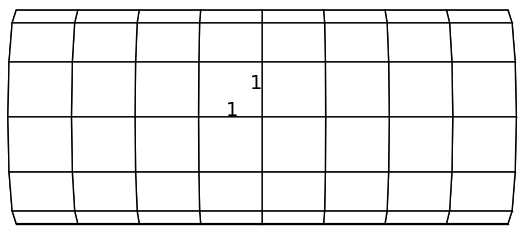}  
	\caption{The surface of a hollow cylinder is divided into $L$ annular rings of $1$ m width, each subdivided into $K$ square sub-areas with $1$ m side length.}
	\label{fig:cylinder}
\end{figure}

Following the method for calculating the capacitance of a cube, we applied the method of moments  to determine the capacitance of a hollow cylinder. In Ref.~\cite{harrington1993field}, the capacitance was estimated by subdividing a 1-meter-long hollow cylinder into ten annular rings. In this article, we not only increased the number of annular rings dividing the hollow cylinder but also further subdivided each ring into square sub-areas. By exploiting its axial symmetry, the surface of the hollow cylinder is divided into \(L\) annular rings with a width of $1$ m, each of which is further subdivided into \( K \) square sub-areas with side lengths of $1$ m, as illustrated in Fig.~\ref{fig:cylinder}. The electric potential at the center of a sub-area generated by its own charge is $3.52549\sigma/(4\pi\varepsilon_0)$ \eqref{eq:inter1}, while contributions from other sub-areas are $\sigma/(4\pi\varepsilon_0d)$ \eqref{eq:inter2}, where $\sigma$ is the charge density and $d$ the center-to-center distance between sub-areas. Thus, the electric potential of an annular ring generated by its own charge can be expressed as a summation:  
\begin{equation}
\sum\limits_{n = 1}^{K - 1} {\frac{\pi \sigma }{{4\pi\varepsilon_0K\sin \frac{{n\pi }}{K}}}}  +  \frac{3.52549\sigma}{4\pi\varepsilon_0}\label{eq:sp}
\end{equation}
Here, \( K \) is also the circumference of the annular ring. The electric potential generated by one annular ring at another can be derived as  
\begin{equation}
\sum\limits_{n = 0}^{K - 1} {\frac{\sigma}{{{{4\pi\varepsilon_0\left( {{d^2} + \frac{{{K^2}}}{{{\pi ^2}}}{{\sin }^2}\left( {\frac{{n\pi }}{K}} \right)} \right)}^{\frac{1}{2}}}}}}\label{eq:mp}
\end{equation}
\( d \) denotes the distance between the centers of the two annular rings.  

Setting the electric potential of every annular ring to $1$ V allows us to solve for the surface charge density $ \sigma $. The capacitance is then obtained directly from the total charge.  For geometrically similar cylindrical configurations, the ratio of capacitance to cylinder radius $R$ remains constant. We computed this capacitance-to-radius ratio for different $L/K$ values and compared the results with Lekner's~\cite{lekner2021electrostatics} numerical calculations, as summarized in Table ~\ref{tab:capa}. 

\begin{table}[h]
	\centering
    \caption{Comparison of Capacitance-to-Radius Ratios $C/R$ [pF/m] for Hollow Cylinders at Different $L/K$ Ratios: Our  Numerical Results vs Lekner's Truncated Matrix Method (6th-Order)~\cite{lekner2021electrostatics}. Lekner's original values are shown in parentheses. $L$: Cylinder Length [m]; $K$: Base Circumference [m]; $R$: Radius ($R=K/2\pi$).}
	\begin{tabular}{ccccc}
		\toprule 
	    $L/K $ & our results ($C/R$) & Lekner ($N=6$) ($C/R$)\\ 	
		\midrule 
		$1000/94248$   & $56.6181$ &   $56.6184$\,($0.508884$) \\ 
		$1000/31416$ & $68.8969$ &   $68.8973$\,($0.619246$) \\  
		$5000/47124$ & $90.5870$ &    $90.5869$\,($0.814191$) \\  
        $5000/31416$ & $101.4889$ &    $101.4888$\,($0.912177$) \\ 
        $5000/23562$ & $111.0306$ &    $111.0305$\,($0.997937$) \\  
        $5000/15708$ & $127.9917$ &    $127.9915$\,($1.150382$) \\ 
        $5000/7854$ & $171.1675$ &    $171.1671$\,($1.538442$)\\ 
        $5000/5236$ & $208.7809$ &    $208.7801$\,($1.876506$)\\ 
        $5000/1571$ & $421.5004$ &    $421.5266$\,($3.788663$)\\ 
        $5000/524$ & $900.2549$ &    $900.6987$\,($8.095440$)\\ 
		\bottomrule 
	\end{tabular} 
	
	\label{tab:capa}
	\vspace{0.5em}
\end{table}

It can be observed that our numerical results for the hollow cylinder capacitance show excellent agreement with Lekner's data. However, slightly larger discrepancies were observed in the $L/K$ ratios of $ 5000/1571$ and $5000/524$, mainly due to the limited computational resources that restricted the maximum value of $L$ to 5000, while the smaller values of $K$ resulted in increased errors.
\begin{figure}[h]
	\centering
	\includegraphics[width=8cm,height=8cm]{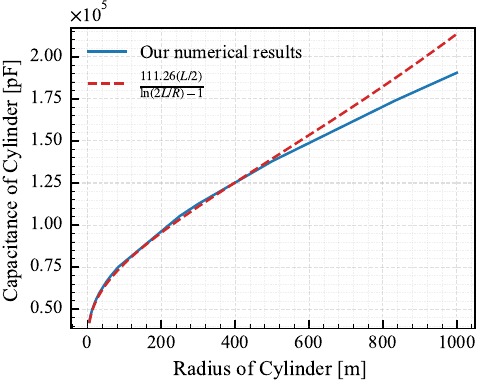}  
	\caption{The capacitance curve of a hollow cylinder versus radius $R$ [m] for length $L = 5000$ [m]. Blue solid curve: our numerical results; red dashed curve: theoretical approximation  $\frac{111.26(L/2)}{\ln(2L/R)-1}$ (valid for $L \gg R$)~\cite{cavendish1879electrical}.
}
	\label{fig:asymp}
\end{figure}

Fig.~\ref{fig:asymp} plots the capacitance of a hollow cylinder as a function of its base radius $R$, with the cylinder length fixed at $L = 5000$ m. The blue solid curve represents our numerical results, and the red dashed curve corresponds to the theoretical approximation $\frac{111.26(L/2)}{\ln(2L/R)-1}$ (valid for $ L \gg R$)~\cite{cavendish1879electrical}. The capacitance increases with $R$, as expected theoretically. Furthermore, as the radius $R$ of the hollow cylinder gradually decreases, the two curves tend to coincide. This confirms the validity of the theoretical approximation formula for the capacitance when the cylinder length $L$ is much larger than the radius $R$. In addition, we observe that when the cylinder radius $R$ exceeds $500$ m (i.e., when $\frac{L}{R} < 10$), the two curves begin to diverge, indicating that the approximation formula $ \frac{111.26(L/2)}{\ln(2L/R)-1} $ yields relatively accurate results for the capacitance of a hollow cylinder when $\frac{L}{R} > 10$.

Our numerical results for the hollow cylinder capacitance are compared with Lekner's calculations and Cavendish's experimental measurements in Table~\ref{tab:cape}. In our computations, the parameters $L$ and $K$ were assigned scaled-up integer values through proportional amplification. Our numerical results show good agreement with those calculated using Eq.\eqref{eq:r1} derived by Lekner, while exhibiting a relatively larger discrepancy when compared to Cavendish's experimental measurements. This discrepancy can be attributed not only to the errors inherent in the theoretical calculations but also to the measurement errors associated with Cavendish's experiment itself. A review of the original notes detailing his experimental procedures reveals that, constrained by the technological limitations of the time, several factors contributed to the experimental uncertainty. These included visual judgment errors in reading the pith-ball electrometer, charge leakage, and imperfect insulation, as documented in his records~\cite{cavendish1879electrical}.

\begin{table}[h]
	\centering
    \caption{Comparison of Hollow Cylinder Capacitance Values [pF] (Numerical Solutions of This Work), Lekner's~\cite{lekner2021electrostatics} 6th-Order Matrix Truncation Solutions, and Cavendish's Experimental Data~\cite{cavendish1879electrical} with Length $L$ and base circumference $K$ in Meters. original values in inches from Cavendish’s measurements are given in parentheses).}
	\begin{tabular}{ccccc}
		\toprule 
	     $L$&$K$ & our results & Lekner ($N=6$)& measured by Cavendish\\ 	
		\midrule 
		$1.3767$ & $0.0583$ & $16.5876$ & $16.5893$ & $16.2609$ ($5.754$) \\  
		$0.9119$ & $0.2019$ & $17.0711$ & $17.0712$ & $17.0804$ ($6.044$)\\  
		\bottomrule 
	\end{tabular} 
	
	\label{tab:cape}
\end{table}
\begin{figure}[h]
	\centering
	\includegraphics[width=8cm,height=8cm]{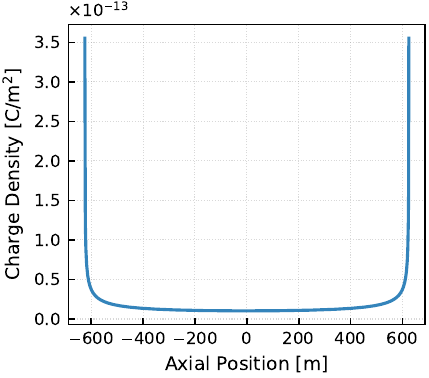}  
	\caption{The Charge Density Distribution Curve of the Hollow Cylinder with   Length \( L= 1250  \) m and  Base Circumference \( K = 3927  \) m ( Diameter \( D = 1250 \) m ).}
	\label{fig:cylinderd}
\end{figure}

As shown in Fig.~\ref{fig:cylinderd}, the charge density distribution of the hollow cylinder exhibits higher values at both ends and lower values in the central region, with sharp variations at the ends and gradual variations in the middle, which is consistent with theoretical expectations.

\section{Discussion and Conclusion}
\label{sec:dis}
This study extends the sub-areas method proposed by D. K. Reitan and T. J. Higgins for calculating  cube capacitance, which later evolved into the method of moments. Reitan and Higgins only employed even divisions of the cube's surface, with a maximum partition scale of $6\times6$ square sub‑areas per face. Building on this, the work of Er‑Wei Bai and Karl E. Lonngren incorporated both even and odd divisions; however, they did not exploit the symmetry among the sub‑areas on the cube's faces. Their maximum partition scale was $20\times20$ sub‑areas per face.

It is worth noting that geometric symmetries among the sub‑areas on the cube's surface still exist even under odd divisions. The present research retains these symmetries fully while being compatible with both even and odd divisions, thereby significantly reducing computational resource requirements. We have increased the maximum partition scale on each face of the cube to $600\times600$ sub‑areas.

Based on the described method, a Python program utilizing a parallel computing architecture was developed and is available from the authors upon request. The computation for a $600\times600$ partition scale takes approximately 1 hour and 43 minutes on a desktop equipped with an 8‑core 11th‑generation Intel i7 processor.

 We find that the computed capacitance of a unit cube (edge length: 1 m) does not converge monotonically with finer discretization. Instead, it first increases, peaks at $73.519014$ pF for $90\times90$ sub-areas per face, and then gradually decreases.  We introduced Thomson's theorem to explain the initial rise in capacitance. In the early stage, as the number of sub-areas on each face of the cube increases, the surface charge distribution gradually approaches electrostatic equilibrium, leading to a reduction in electrostatic energy. According to Eq.\eqref{eq:c8}, the capacitance correspondingly increases and converges toward the theoretical value. However, upon further refinement, errors inherent in the discretization scheme itself become dominant, causing the calculated capacitance to decrease. Through a detailed analysis based on rigorous integral calculations, we find that the potential approximation formula \eqref{eq:inter2} systematically overestimates the actual potential. This results in an underestimated charge density from the inversion process, and consequently leads to an underestimated capacitance. This error becomes dominant under fine discretization, posing a challenge to the method of moments in such problems: even with an unlimited increase in sub-areas, discretization‑induced errors may still prevent the numerical results from converging to the exact value. Finally, we have visualized the charge density distribution on the surface of the cube, as presented in  Fig.~\ref{fig:Charge}. This distribution pattern is consistent with the predictions of electrostatic theory.

Subsequently, this method was applied to calculate the capacitance of a hollow cylinder. The cylinder was divided into $L$ annular rings ($1$ m in width), each subdivided into $K$ square sub-areas ($1$ m side length). We derived analytical expressions for self-potential (Eq.\eqref{eq:sp}) and mutual-potential (Eq.\eqref{eq:mp}) between ring elements. For various $L/K$ ratios, our computed capacitances show excellent agreement with Lekner's numerical solutions and are also consistent with Cavendish's experimental measurements.   In addition, the capacitance of a hollow cylinder versus its radius $R$ is shown in Fig.~\ref{fig:asymp} for a fixed length $L = 5000$ m. Comparison of the numerical results with the theoretical approximation demonstrates that the expression $\frac{111.26(L/2)}{\ln(2L/R)-1}$ provides an accurate description of the capacitance when the aspect ratio satisfies $\frac{L}{R} > 10$. We also present the charge density distribution along the axial direction for a hollow cylinder with an aspect ratio of $1$, as shown in Fig.~\ref{fig:cylinderd}. Its variation trend is consistent with the predictions of electrostatic theory.

A comparison of our method with those of Refs.~\cite{de2026capacitance,lekner2021electrostatics} is instructive. The Chebyshev collocation scheme of de~Sousa~\cite{de2026capacitance} is a high-order spectral method that analytically factors out the edge singularity, producing extremely accurate benchmark values with very few degrees of freedom. Lekner’s~\cite{lekner2021electrostatics} truncated-matrix method is likewise a semi-analytical technique tailored to the hollow cylinder. Both are highly efficient for the high-accuracy computation of the capacitance of hollow cylinders with arbitrary aspect ratios. In contrast, the method of moments  employed here is a flexible, generic discretization technique. For the hollow cylinder, axial symmetry was fully exploited to classify sub-areas and reduce the number of unknowns, and parallel computing made the matrix assembly and solution very efficient. As an illustration, with the Python code developed in this work for the hollow cylinder capacitance, the slowest case ($L/K=5000:47124$) takes only $891$ seconds using parallel computing on a desktop with an 8-core 11th-generation Intel i7 processor, while the fastest case ($L/K=1000:31416$) takes just $7$ seconds, demonstrating excellent numerical performance. Thus, the advantages of our approach lie in its conceptual simplicity, clear physical picture, and ease of implementation. By exploiting symmetry and parallel computing to handle a very large number of sub-areas, it yields highly accurate capacitance values for hollow cylinders of arbitrary aspect ratios.

\begin{acknowledgments}
\noindent
We would like to express our gratitude to Zhidao Bu for his significant assistance in literature retrieval. 
\end{acknowledgments}
\section*{Declarations}
H.G. is supported by the Fujian Provincial Education and Scientific Research Program for Young and Middle-aged Teachers (Science \& Technology Category) under Grant No. JAT231183. P.Y. is supported by the Fujian Provincial Education and Scientific Research Program for Young and Middle-aged Teachers (Science \& Technology Category) under Grant No. JZ230071.

\appendix

\section{Electric Potential at the Center and in the Exterior Region of a Unit Square}
\label{appendix:ElectricPotential}
\begin{figure}[h]
	\centering
	\includegraphics[width=5cm,height=5cm]{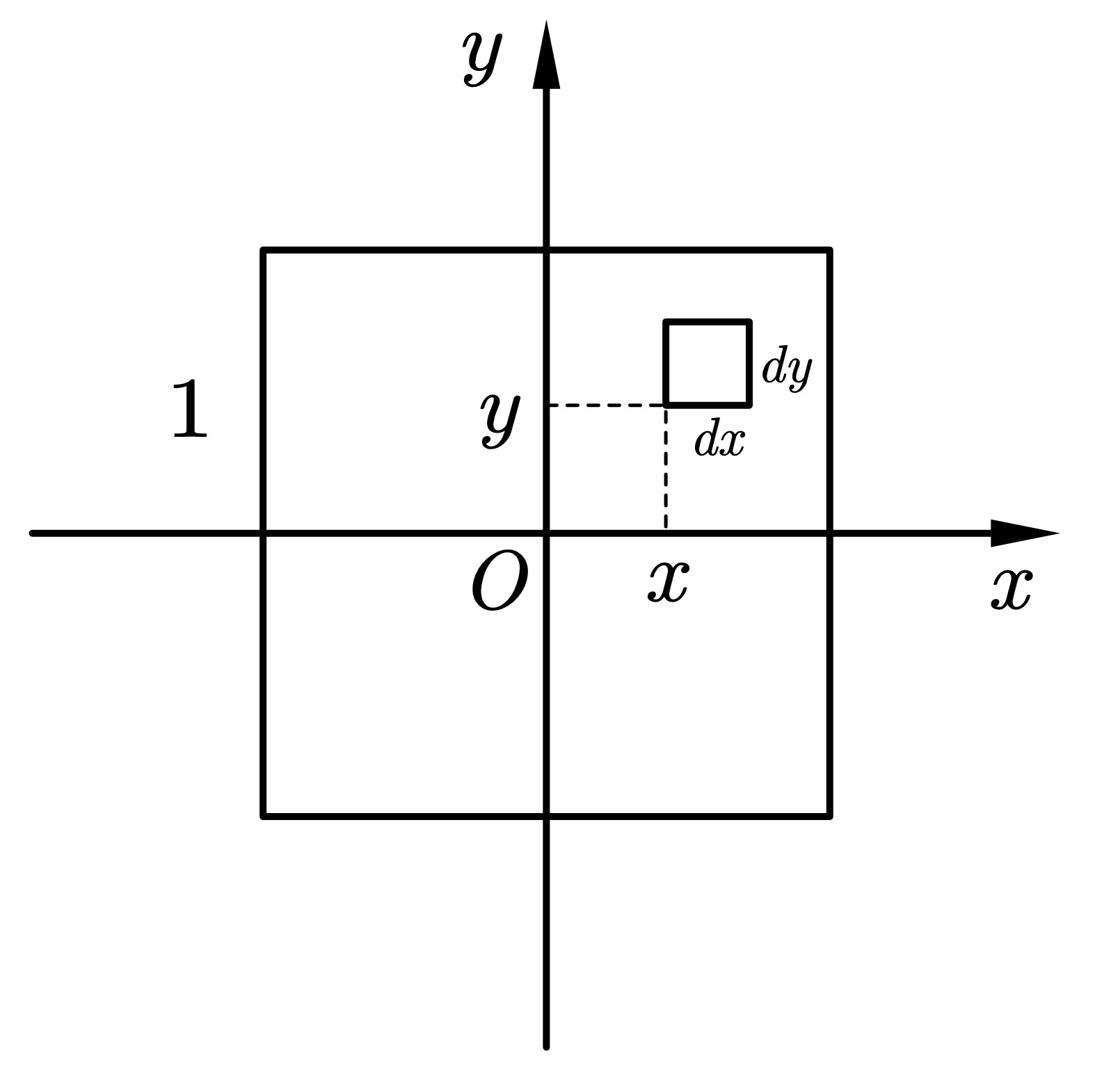}  
	\caption{Schematic diagram of the electric potential calculation at the center of a unit square. \newline\footnotesize Reproduced from [Daniel Kinseth Reitan and Thomas James Higgins. Calculation of the electrical capacitance of a cube. Journal of applied Physics, 22(2):223–226, 1951.], with the permission of AIP Publishing.}
	\label{fig:CubeA}
\end{figure}

Assuming a square with a side length of $1$ m as shown in Fig.~\ref{fig:CubeA}, where the charge density $\sigma$ is uniformly distributed, the electric potential at the center of the square can be expressed as~\cite{dwight1988tables}
\begin{equation}
\begin{aligned}
\phi &= \frac{4\sigma}{4\pi\varepsilon_0} \int_{0}^{\frac{1}{2}} \int_{0}^{\frac{1}{2}} \frac{dxdy}{(x^2 + y^2)^{\frac{1}{2}}}
   = \frac{4\sigma}{4\pi\varepsilon_0}\int_{0}^{\frac{1}{2}} \sinh^{-1}\frac{1}{2y} dy\\
  &= \frac{2\sigma}{4\pi\varepsilon_0} \int_{1}^{\infty} u^{-2} \sinh^{-1}u du
   = \frac{4\sigma}{4\pi\varepsilon_0} \ln[1 + (2)^{\frac{1}{2}}]\\
   &= \frac{3.52549\sigma}{4\pi\varepsilon_0}\label{eq:inter1}
\end{aligned}
\end{equation}

\begin{figure}[htbp]
	\centering
	\includegraphics[width=7cm,height=7cm]{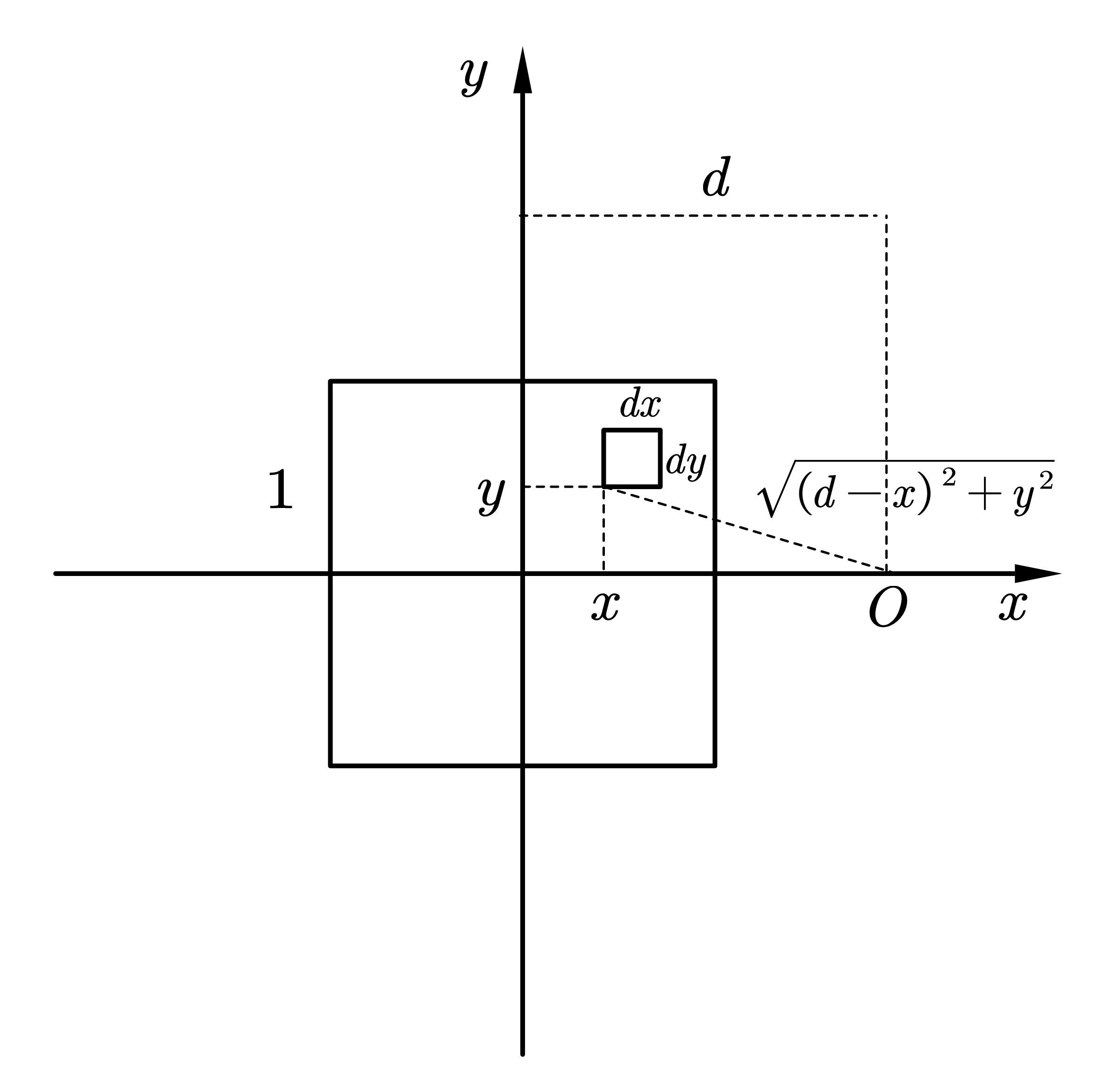}  
	\caption{Schematic diagram of the electric potential calculation at a distance \( d \) (\( d > 1 \)) from the center of a unit square. \newline\footnotesize Reproduced from [Daniel Kinseth Reitan and Thomas James Higgins. Calculation of the electrical capacitance of a cube. Journal of applied Physics, 22(2):223–226, 1951.], with the permission of AIP Publishing.}
	\label{fig:CubeB}
\end{figure}
When point $O$ is located outside the square at a distance \( d \) from its center, as shown in Fig.~\ref{fig:CubeB}, the electric potential can be expressed as~\cite{dwight1988tables}
\begin{equation}
\phi=\frac{\sigma}{4\pi\varepsilon_0}\int_{-\frac{1}{2}}^{\frac{1}{2}}\int_{-\frac{1}{2}}^{\frac{1}{2}}\frac{dxdy}{\sqrt{(d - x)^2 + y^2}} \sim \frac{\sigma}{4\pi\varepsilon_0d}\label{eq:inter2}
\end{equation}

\section{Coefficient Matrix $\mathbf{P}$}
\label{app:coefficient_matrix}

The coefficient matrix $\mathbf{P}$ from the main matrix Eq.\eqref{eq:matrix_equation_main} is given below. The element $P_{ij}$ in the $i$-th row and $j$-th column corresponds to the coefficient of $\sigma_j$ in the equation for $\phi_i$.

\begin{table}[htbp]
  \centering
  \caption[Complete coefficient matrix $\mathbf{P}$]{%
    Complete elements of the $10 \times 10$ coefficient matrix $\mathbf{P}$.  Note: The elements of this coefficient matrix need to be divided by $4\pi\epsilon_0$.}
  \footnotesize
  \begin{tabular}{ccccccccccc}
    \toprule
    $i \setminus j$ & $1$ & $2$ & $3$ & $4$ & $5$ & $6$ & $7$ & $8$ & $9$ & $10$ \\
   \midrule
    $1$  & $9.052$ & $10.581$ & $8.743$ & $4.181$ & $4.760$ & $8.685$ & $4.222$ & $4.232$ & $4.184$ & $1.040$ \\
    $2$  & $5.290$ & $13.204$ & $9.773$ & $4.553$ & $5.235$ & $9.520$ & $4.597$ & $4.600$ & $4.539$ & $1.127$ \\
    $3$  & $4.371$ & $9.773$ & $13.117$ & $5.398$ & $4.959$ & $10.215$ & $5.066$ & $4.881$ & $4.845$ & $1.199$ \\
    $4$  & $4.181$ & $9.106$ & $10.796$ & $8.515$ & $4.778$ & $10.223$ & $5.473$ & $4.975$ & $4.981$ & $1.228$ \\
    $5$  & $4.760$ & $10.470$ & $9.919$ & $4.778$ & $7.851$ & $10.748$ & $5.015$ & $5.155$ & $5.017$ & $1.244$ \\
    $6$  & $4.342$ & $9.520$ & $10.215$ & $5.111$ & $5.374$ & $13.384$ & $5.637$ & $5.726$ & $5.549$ & $1.365$ \\
    $7$  & $4.222$ & $9.195$ & $10.132$ & $5.473$ & $5.015$ & $11.274$ & $8.200$ & $5.744$ & $5.938$ & $1.428$ \\
    $8$  & $4.232$ & $9.199$ & $9.762$ & $4.975$ & $5.155$ & $11.451$ & $5.744$ & $8.603$ & $6.635$ & $1.646$ \\
    $9$  & $4.184$ & $9.078$ & $9.690$ & $4.981$ & $5.017$ & $11.098$ & $5.938$ & $6.635$ & $9.204$ & $1.945$ \\
    $10$ & $4.161$ & $9.015$ & $9.590$ & $4.914$ & $4.977$ & $10.922$ & $5.712$ & $6.582$ & $7.782$ & $4.476$ \\
   \bottomrule
  \end{tabular}
  
  \label{tab:p_matrix_normal} 
\end{table}

\section{Capacitance of a Unit Cube}
\label{appendix:Capacitancecube}
\begin{table}[h]
	\centering
    \caption{Calculated Capacitance Values for a Unit Cube (edge length: $1$ m) with Each Face Divided into Varying Numbers of Sub-areas.}
\begin{tabular}{cc}
\toprule
Number of Sub-areas & Results $[\mathrm{pF}]$\\
\midrule
 $7\times7$   & $73.074699$   \\
 $10\times10$ & $73.280263$   \\
 $20\times20$ & $73.458128$  \\
 $30\times30$ & $73.496007$   \\
 $40\times40$ & $73.509142$  \\
 $50\times50$ & $73.514687$ \\
 $60\times60$ & $73.517232$   \\
 $70\times70$ & $73.518406$   \\
 $80\times80$ & $73.518892$ \\
 $85\times85$ & $73.518986$   \\
 $89\times89$ & $73.519013$   \\
 $90\times90$ & $73.519014$   \\
 $91\times91$ & $73.519013$  \\
 $95\times95$ & $73.518991$ \\
 $100\times100$ & $73.518934$  \\
 $200\times200$& $73.516239$ \\
 $300\times300$ & $73.514237$   \\
 $400\times400$ & $73.512946$\\
 $500\times500$ & $73.512054$\\
 $600\times600$ & $73.511405$\\
\bottomrule
\end{tabular}

\label{tab:resultsB}
\end{table}

\section{Discretization Error Analysis}
\label{appendix:Error}
Taking the center of a cube face as the origin (e.g., the center of the sub-area marked $10$ in Fig.~\ref{fig:Cube}), the electric potential it produces at other target sub-areas on the cube surfaces is calculated. The difference between the results from the full integral expression and the approximate formula (both given in formula \eqref{eq:inter2}) is analyzed. Three cases are examined separately:
\begin{itemize}[leftmargin=0pt]
\item
When the two sub-areas are coplanar, the potential difference $ \Delta \phi$ is given by the following expression, where $x_1$ and $y_1$ are the coordinates of the center of the target sub-area. We define $\mathbf{r} = (x_1, y_1, 0)$ and $\mathbf{r}' = (x, y, 0)$.
\begin{equation}    
\begin{aligned}
\Delta \phi &= \frac{\sigma}{4\pi\varepsilon_0}\Bigg[\frac{1}{\sqrt{x_1^2 + y_1^2}} - \int_{-\frac{1}{2}}^{\frac{1}{2}} \int_{-\frac{1}{2}}^{\frac{1}{2}} \frac{dx\, dy}{\sqrt{(x_1 - x)^2 + (y_1 - y)^2}}\Bigg] \\
&=\frac{\sigma}{4\pi\varepsilon_0}\Bigg(\frac{1}{r}-\int_{-1/2}^{1/2}\int_{-1/2}^{1/2} \frac{dx\, dy}{|\mathbf{r} - \mathbf{r}'|}\Bigg),\quad r = \sqrt{x_1^2 + y_1^2}\\
&=\frac{\sigma}{4\pi\varepsilon_0}\Bigg\{\frac{1}{r}-\int_{-1/2}^{1/2}\int_{-1/2}^{1/2} dxdy\Bigg[\frac{1}{r} + \frac{\mathbf{r}\cdot\mathbf{r}'}{r^3} \\
&\quad + \frac{1}{2}\Bigg(\frac{3(\mathbf{r}\cdot\mathbf{r}')^2}{r^5} - \frac{|\mathbf{r}'|^2}{r^3}\Bigg) + \cdots\Bigg]\Bigg\} \\
&=\frac{\sigma}{4\pi\varepsilon_0}\Bigg\{\frac{1}{r}-\Bigg[\frac{1}{r} + \frac{1}{24r^3} + O\Bigg(\frac{1}{r^5}\Bigg)\Bigg]\Bigg\}\\
&=\frac{\sigma}{4\pi\varepsilon_0}\Bigg[ -\frac{1}{24r^3} + O\Bigg(\frac{1}{r^5}\Bigg)\Bigg] < 0
\end{aligned}
\end{equation}

It can be observed that when two sub-areas are coplanar, the potential value yielded by the approximate formula \eqref{eq:inter2} is consistently lower than that obtained via two-dimensional integration, regardless of their separation distance.

\item  When the two sub-areas are on opposite faces, the potential difference $ \Delta \phi$ is given by the following expression, where $z_1$ equals the edge length of the cube. Moreover, since the coordinate origin is located at the center of the face, we have $x_1, y_1 < z_1/2$. Hence, for this configuration, the potential calculated using the approximate formula \eqref{eq:inter2} is greater than that obtained through integration (Note that here we have only considered the potential generated by a sub-area at the center of a cube face on its opposite face).
\begin{equation}
\begin{split}
\Delta \phi &= \frac{\sigma}{4\pi\varepsilon_0}
\Bigg[ \frac{1}{\sqrt{x_1^2 + y_1^2 + z_1^2}} \\
&\quad - \int_{-1/2}^{1/2}\int_{-1/2}^{1/2} 
\frac{dx\, dy}{\sqrt{(x_1 - x)^2 + (y_1 - y)^2 + z_1^2}} \Bigg] \\
&\approx \frac{\sigma}{4\pi\varepsilon_0}
\Bigg[ \frac{1}{r_0} - \bigg( \frac{1}{r_0} + \frac{x_1^2 + y_1^2}{8 r_0^5} 
- \frac{1}{12 r_0^3} \bigg) \Bigg]  \\
&\approx \frac{\sigma}{4\pi\varepsilon_0} 
\cdot \frac{2z_1^2 - (x_1^2 + y_1^2)}{24 r_0^5} > 0,\,\,\, r_0 = \sqrt{x_1^2 + y_1^2 + z_1^2}
\end{split}
\end{equation}

\item  When the two sub-areas are on adjacent faces, the potential difference $\Delta \phi$ is given by the following expression, where $x_1$ equals half the edge length of the cube.
\begin{equation}
\begin{split}
\Delta \phi &= \frac{\sigma}{4\pi\varepsilon_0}
\Bigg[ \frac{1}{\sqrt{x_1^2 + y_1^2 + z_1^2}} \\
&\quad - \int_{-1/2}^{1/2}\int_{-1/2}^{1/2} 
\frac{dx\, dy}{\sqrt{(x_1 - x)^2 + (y_1 - y)^2 + z_1^2}} \Bigg] \\
&\approx \frac{\sigma}{4\pi\varepsilon_0}
\Bigg[ \frac{1}{r_0} - \bigg( \frac{1}{r_0} + \frac{x_1^2 + y_1^2}{8 r_0^5} 
- \frac{1}{12 r_0^3} \bigg) \Bigg]  \\
&\approx \frac{\sigma}{4\pi\varepsilon_0} 
\cdot \frac{2z_1^2 - (x_1^2 + y_1^2)}{24 r_0^5},\,\,\,\, r_0 = \sqrt{x_1^2 + y_1^2 + z_1^2}
\end{split}
\end{equation}
Here, the sign of the electric potential difference $\Delta \phi$ depends on the values of $x_1$, $y_1$, and $z_1$. We performed numerical calculations to determine the proportion of cases where $\Delta \phi$ takes positive values when the cube is divided into different numbers of sub-areas. The results are presented in Fig.~\ref{fig:error}.
\begin{figure}[h]
	\centering
	\includegraphics[width=8cm,height=8cm]{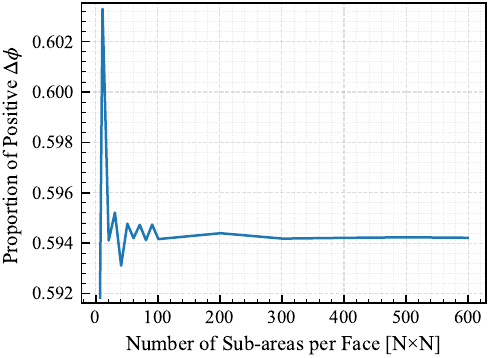}  
	\caption{Proportion of Positive $\Delta \phi$ with Each Face Divided into N$\times$ N Sub-areas (N: X-Axis).}
	\label{fig:error}
\end{figure}
\end{itemize}
From Fig.~\ref{fig:error}, it can be observed that the proportion of positive $\Delta \phi$ values remains essentially constant at approximately 59\%, showing no significant dependence on the number of sub-areas used in the discretization. However, as the total number of sub-areas, $6\times N^2$, increases, the difference between the number of sub-areas where $\Delta \phi$  takes positive values and those where it takes negative values also increases, meaning that the overestimation effect on the electric potential is enhanced. Considering that the cube possesses four adjacent faces and by synthesizing the two scenarios discussed above, we can draw a general conclusion: the electric potential calculated using the approximate formula \eqref{eq:inter2} is systematically higher than the value obtained from the exact integral expression.

Since the potential for each sub-area is fixed at $1$ V, this systematic overestimation leads to an underestimation of the computed charge density, which in turn results in a reduced capacitance value. As the number of sub-areas per cube face increases, the positive effect of finer discretization on enhancing the capacitance calculation becomes negligible. Beyond this point, the error inherent in the adopted potential approximation emerges as the dominant factor governing the result.

\newpage

\bibliography{Refs}

\bibliographystyle{unsrtnat}

\end{document}